\documentclass[12pt,a4]{article}
\input{epsf.tex}
\begin{document}
\begin{center}
{\Large \bf
ESFRAD. FORTRAN code for
calculation of QED corrections
to polarized ep-scattering \\ by
the electron structure function
method }
\vspace{4 mm}

A. Afanasev$^{a)}$, I. Akushevich$^{b)}$, A. Ilyichev$^{c)}$, N.
Merenkov$^{d)}$

\vspace{4 mm}
$^{(a)}$Jefferson Lab, Newport News, VA 23606, USA\\
$^{(b)}$Duke University,
Durham, NC 27708, USA \\
$^{(c)}$ National Center of Particle and High Energy Physics,
220040 Minsk, Belarus \\
$^{(d)}$ NSC ''Kharkov Institute of Physics and Technology'',\\
61108 Kharkov, Ukraine
\end{center}

\begin{abstract}
The main features of the electron structure function method for
calculations of the higher order QED radiative effects
to polarized deep-inelastic $ep$-scattering are presented.
A new FORTRAN code ESFRAD based on this method was developed.
A detailed quantitative comparison between the results of ESFRAD
and  other methods implemented in the codes POLRAD and  RADGEN
for calculation of the higher order radiative corrections
is performed.
\end{abstract}

\section{Introduction}
The observed cross section in the electron--proton deep inelastic scattering
\begin{equation}
e^-(k_1)+p(p_1)\to
e^-(k_2)+X(p_x)
\end{equation}
($k_1^2=k_2^2=m^2$)
is affected by electromagnetic effects caused by real and virtual
photon emission and $e^+e^-$ -- pair production, which cannot
be removed with any experimental cuts.
Therefore contributions of these effects called radiative
corrections (RC), to different observables has to be calculated
theoretically and then subtracted from experimental data
(see \cite{rev} for details).

In the present report the model--independent RC are considered.
Such RC include contributions due to radiation of real but not
observed photons and electron--positron pairs as well as loop
corrections in the lepton part of interaction. The model--dependent
RC arise from electromagnetic effects in the hadron part of interaction.
Such RC are described by the box--type and proton vertex diagrams
and real photon radiation by the proton.
Their calculation requires additional assumptions about the hadron
structure and, therefore, they have additional theoretical
uncertainties. The model--dependent RC are the subject of separate
theoretical studies (see, e.g,
Ref.~\cite{mdep1, mdep2, mdep3, mdep4} for the elastic
$ep$-scattering case).

The structure of model--independent RC in ultrarelativistic approximation has a
simple form. The contribution of virtual and real particles to the observed cross
section in the $n$-th order of perturbation theory  can be presented as the following sum,
\begin{equation}
\frac {d\sigma _{n}^{RC}}{dQ^2dy}=
\left(\frac {\alpha}{\pi}\right)^n\sum _{i=0}^nC_{ni}(Q^2,y)L^i
+{\cal O}(\frac {m^2}{Q^2}),\;
\label{nr}
\end{equation}
where $Q^2=-(k_1-k_2)^2$, $y=(k_1p_1-k_2p_1)/k_1p_1$,
$L=\log Q^2/m^2$ and the coefficients $C_{ni}(Q^2,y)$
are independent on the electron mass. Each of the leading terms
$C_{ni}(Q^2,y)L^i$ describes along  with elastic
(i.e., without no emitted photons) and inelastic events with
the number of hard collinear photons up to $n$. Since at modern
accelerator energies the quantity $L$ is about 10 - 20, the main
contribution to RC (2) gives so-called leading order (LO)
terms which are proportional to $(\alpha L/\pi )^n$ and all the
real photons as well as $e^+e^-$-pairs in this case are collinear
of the parental electrons.
Note that the contribution $\sim (\alpha /\pi )^nL^{n-1}$ is
usually called  next-to-leading order (NLO).

At present only the lowest order radiated corrections ($n=1$) have
been calculated completely for inelastic \cite{ash}
and elastic lepton-hadron scattering \cite{el} using the covariant
approach developed in \cite{BSh}. The formalism of \cite{ash}
was used later to develop a FORTRAN code POLRAD \cite{pol}
and a Monte Carlo generator RADGEN \cite{rad} for simulation of
radiative events in deep--inelastic $ep$--scattering.

Unfortunately, evaluation of the higher order effects with $n \geq 2$
were estimated not so precise as lowest ones because of the extremely
complicated integration over the unobserved real photon phase space.
Firstly, the soft photon approximation was used by
by D. Yennie, S. Frauchi and H. Suura \cite{YFS} to describe the multi--photon
radiation.
Then the calculation of RC within the electron structure function (ESF) method
was proposed by Fadin and Kuraev \cite{KF} who applied the Drell--Yan
representation
for the electron--positron annihilation cross section. According to this
method, the radiatively corrected observed cross section can be written as a
convolution of two ESF with its hard part. Within the leading accuracy the hard
part coincides with the Born cross section.

Further development of the ESF method
was done in \cite{KMF}. It consists in modification of the hard part
of the cross section taking into account the possibility of one non--collinear
hard photon radiation. Such modification allows to go beyond the leading
approximation and to compute partly
NLO contributions into RC of the
order $(\alpha/\pi)^nL^{(n-1)}$
\begin{equation}
\frac {d\sigma _{LL}^{RC}}{dQ^2dy}=\sum _{n=1}^{\infty}
\left(\frac {\alpha }
{\pi}L\right )^n C_{nn}(Q^2,y)+\left(\frac {\alpha}
{\pi}\right)^n L^{(n-1)} C_{nn-1}(Q^2,y),
\end{equation}
while the NLO contribution of the order $\alpha/\pi$ is calculated
exactly. There exists another source of the NLO terms like
the second one in the right--hand side of Eq.(3), which could not be
evaluated in \cite{KMF}. Such terms arise due
to the improvement of ESFs by including their
non-leading pieces. The respective calculations are in progress now.

The second order correction ($\sim (\alpha
L/\pi)^2$) for unpolarized and polarized inelastic
lepton-hadron scattering was evaluated numerically
in \cite{KurSpi} and \cite{LO}, respectively, just  the ESF method.
In this report we analyze the results of the new FORTRAN code ESFRAD
\cite{ESF}
based on approach developed in \cite{KMF} for both unpolarized and polarized
deep-inelastic scattering
and compare it with the codes POLRAD and RADGEN that treat the
higher--order QED effects differently.

\section{Master Formula}

\begin{figure}[t]
\unitlength 1mm
\vspace*{35mm}
\begin{picture}(20,20)
\put(35,-54){
\epsfxsize=9cm
\epsfysize=9cm
\epsfbox{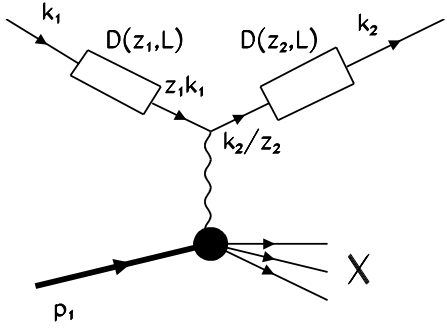}
}
\end{picture}
\vspace*{-1cm}
\caption{Diagram for the corrected cross section within the ESF method }
\label{ME}
\end{figure}

A master formula for the
observed cross section within the ESF method is derived according to
Fig.~\ref{ME} that should be considered as a diagram for the cross section
(not for the amplitude),
\begin{equation}
\frac{d^2\sigma^{obs}(k_1,k_2)}{dQ^2dy}=
\int\limits^1_{z_{1m}}dz_1
\int\limits^1_{z_{2m}}dz_2
D(z_1,L)\frac 1{z_2^2}
D(z_2,L)
\frac{d\sigma^{hard}(\tilde{k}_1,\tilde{k}_2)}
{d\tilde{Q}^2d\tilde{y}},\;
\end{equation}
where
the physical meaning of integration variables $z_1$ and $z_2$ can be
understood in Fig.1. The hard cross section in the integrand depends on
so--called shifted variables which are defined as follows
\begin{equation}
\tilde{k}_1=z_1k_1,\;\tilde{k}_2=\frac{k_2}{z_2},\;
\tilde{Q}^2=\frac{z_1}{z_2}Q^2,\;
\tilde{y}=1-\frac{1-y}{z_1z_2}.
\end{equation}
The lowest limits of integration can be found as a solution of the
standard $\pi$-production threshold inequality for  DIS process:
$(p_1+\tilde{k}_1-\tilde{k}_2)^2>(M+m_{\pi})^2$ that leads to
\begin{equation}
z_{2m}=\frac {1-y+xyz_1}{z_1-z_{th}},\;
z_{1m}=\frac {1-y+z_{th}}{1-xy},\;
z_{th}=\frac{m_{\pi }(m_{\pi}+2M)}{2p_1k_1}
\end{equation}

The electron structure function $D(z_1,L)$ gives the energy
fraction $(z_1)$ distribution of electrons with virtuality
up to $Q^2$ which is created by the initial electron,
whereas $D(z_2,L)$ is in fact the fragmentation function
of the heavy intermediate electron into real one with the energy fraction $z_2.$
Function $D(z,L)$ absorbs all the leading contributions connected with
real and virtual photon emission and pair productions
\begin{equation}
D(z,L)=D^{\gamma}(z,L)+D^{e^+e^-}_N(z,L)+D^{e^+e^-}_S(z,L),\;
\int \limits_0^1dz[D^{\gamma}(z,L)+D_N^{e^+e^-}(z,L)] =1.
\end{equation}
Roughly speaking, the contribution to the function $D^{\gamma}$ comes
from all the graphs with radiation
of real and virtual photons, which is represented here in symbolic
form as follows
\vspace*{-50mm}
\begin{figure}[h!]
\unitlength 1mm
\begin{picture}(20,20)
\put(0,-80){
\epsfxsize=9cm
\epsfysize=8cm
\epsfbox{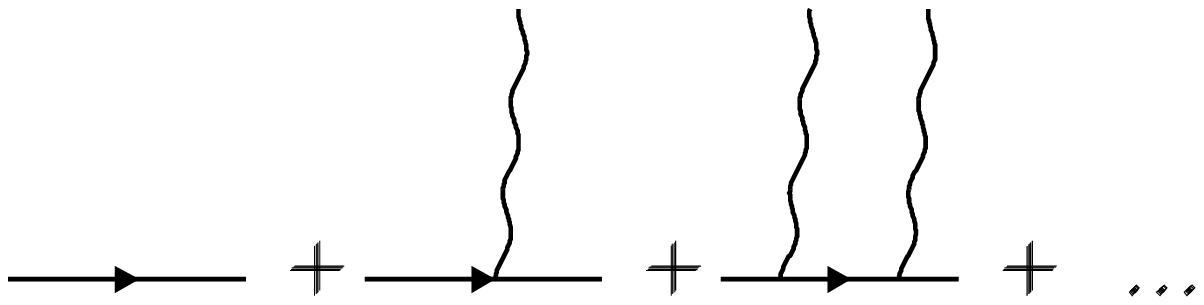}
}
\end{picture}
\end{figure}
\vspace*{0.5cm}

\newpage
\noindent
A nonsinglet part of $D^{e^+e^-}$ is described in the same
graphs but with attached $e^+e^-$-pairs to  the photon lines.
\vspace*{-0.5cm}
\begin{figure}[h!]
\unitlength 1mm
\begin{picture}(20,20)
\put(10,-34){
\epsfxsize=7cm
\epsfysize=5cm
\epsfbox{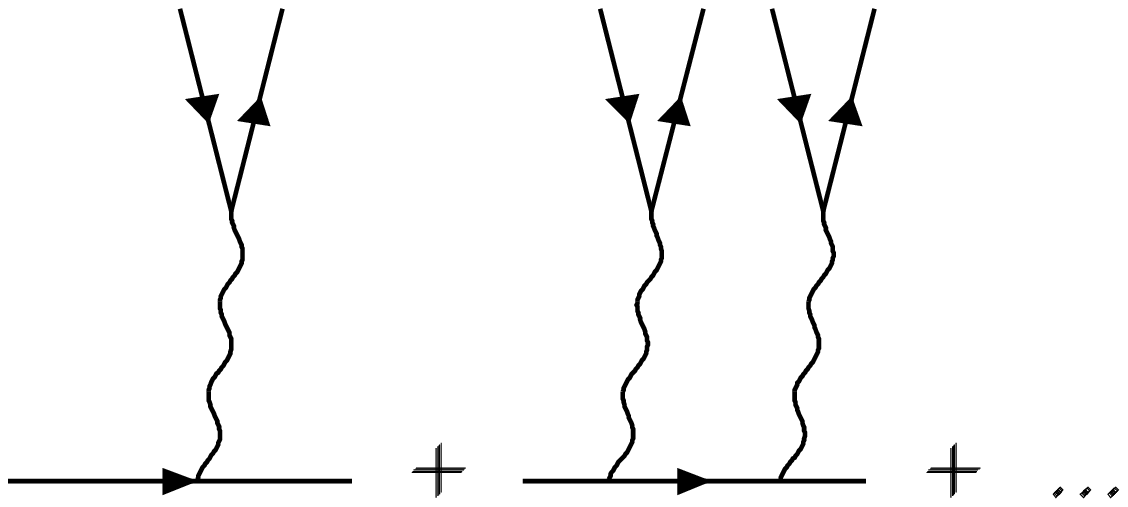}
}
\end{picture}
\end{figure}
\vspace*{0.5cm}

\noindent
The graphs that have at least  one intermediate photon which
propagates along the parental electron line contribute
to the singlet part of $D^{e^+e^-}$.
\vspace*{-20mm}
\begin{figure}[h!]
\unitlength 1mm
\begin{picture}(40,40)
\put(0,-25){
\epsfxsize=8cm
\epsfysize=5cm
\epsfbox{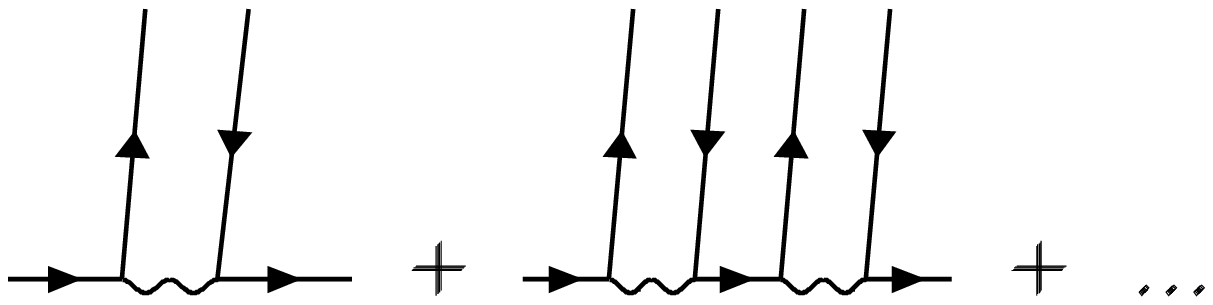}
}
\end{picture}
\end{figure}
\vspace*{-5mm}

\noindent
For details about the $D$-function see, for example, \cite{Jad}.

The hard cross section in the master formula consists of the sum
of the Born cross section and NLO part of lowest order RC.
It can be written as
\begin{eqnarray}
\frac{d\sigma^{hard}(k_1,k_2)}
{dQ^2dy}&=&\frac 1{[1-\alpha /\pi \Pi(Q^2)]^2}
\left[\frac{d\sigma_0(k_1,k_2)}
{dQ^2dy}+
\left(\frac {\alpha}{\pi}\right)C_{10}(Q^2,y)\right],
\end{eqnarray}
where $\Pi(Q^2)$ is an intermediate photon self energy and
the coefficient
$C_{10}(Q^2,y)$ enters Eq.(2).

The explicit expressions for the electron structure functions,
used in ESFRAD  as
well as the more  detailed description of the hard cross section
can be found in \cite{KMF,ESF}.

\section{Numerical Results}

For the numerical analysis performed within experiments on
fixed proton targets at JLab
($E_{beam}=4$ GeV)
and HERMES
($E_{beam}=27.5$ GeV), NMC parametrization
for the unpolarized hadron structure function $F_2$ is used.
The polarized hadron structure functions $g_1$ and
$g_2$ are expressed by the asymmetries
\begin{equation}
A_1=a+x^b(1-\exp (cx)),\;
\label{a1}
\end{equation}
with $a=1.90202\times 10^{-2}$, $b=-1.16312\times 10^{-3}$,
$c=-1.8451$ and
\begin{equation}
A_2=\frac{0.53Mx}{\sqrt{Q^2}}
\label{a2}
\end{equation}
in a standard way.

\begin{figure}[t!]
\unitlength 1mm
\vspace*{1cm}
\begin{tabular}{cc}
\begin{picture}(60,60)
\put(-10,-10){
\epsfxsize=8cm
\epsfysize=5cm
\epsfbox{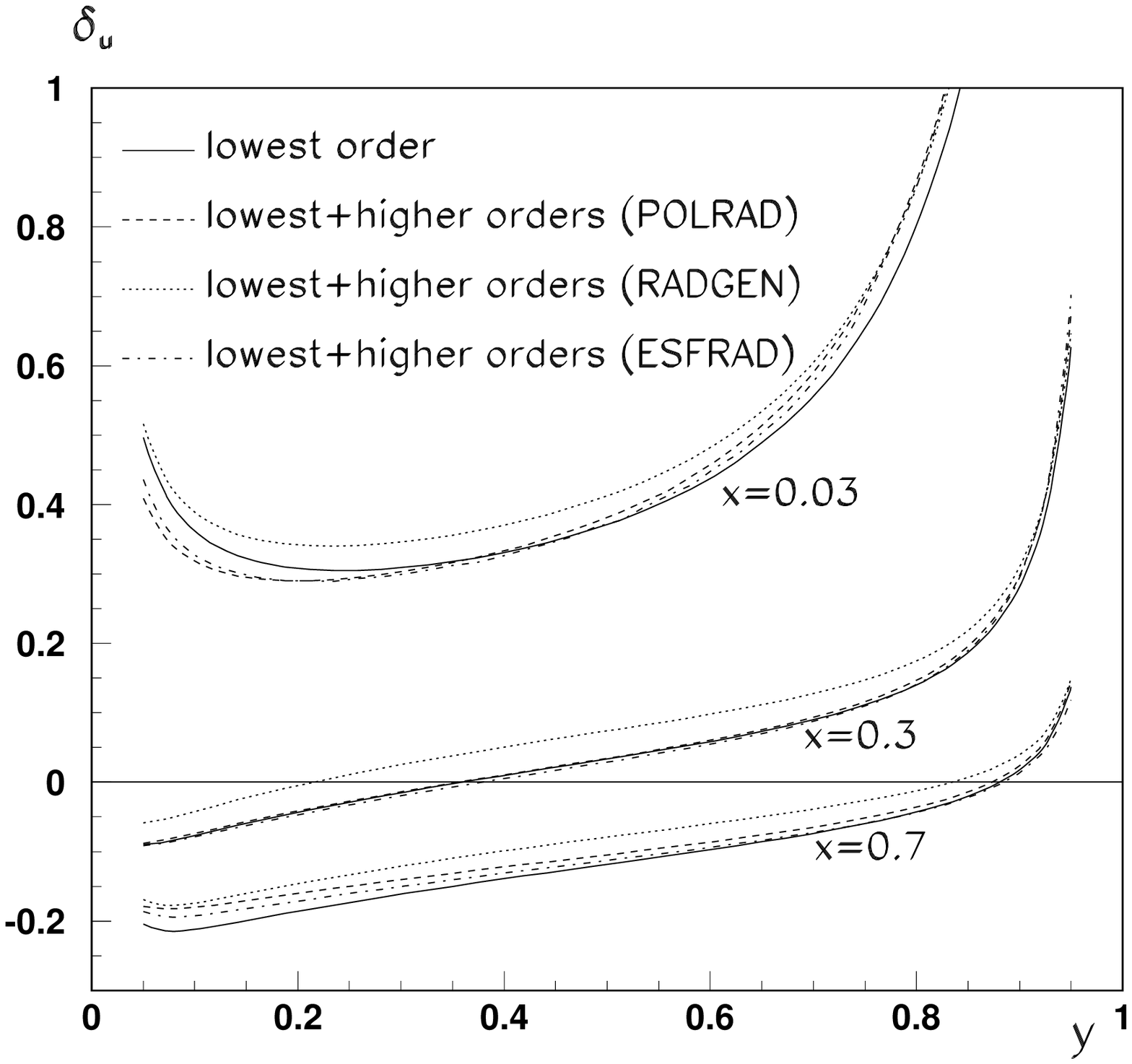}
}
\put(30,-3) {\makebox(0,0){$a)$}}
\end{picture}
&
\begin{picture}(60,60)
\put(10,-10){
\epsfxsize=8cm
\epsfysize=5cm
\epsfbox{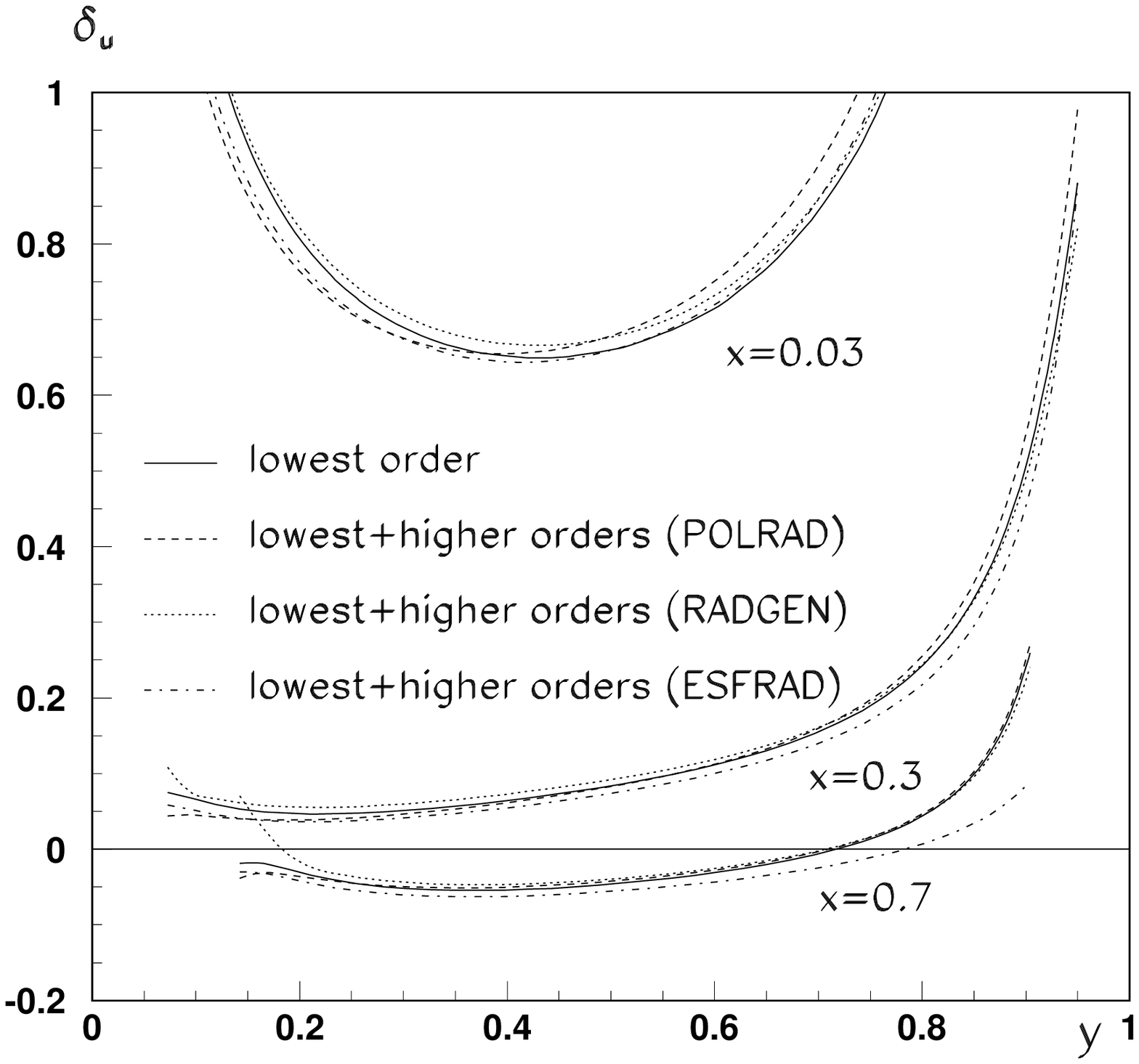}
}
\put(50,-3) {\makebox(0,0){$b)$}}
\end{picture}
\end{tabular}
\vspace*{5mm}
\caption{Relative radiative corrections to the unpolarized cross
section defined by formulae (\ref{d})
for a) HERMES ($E_{beam}=27.57$ GeV) and b) JLab ($E_{beam}=4$ GeV)
kinematic conditions}
\label{du}
\end{figure}

The $y$-dependences of the relative RC to unpolarized part of the
cross section
\begin{equation}
\delta _u=\frac{d\sigma^{obs}/dQ^2dy-d\sigma^{0}/dQ^2dy}
{d\sigma^{0}/dQ^2dy},
\label{d}
\end{equation}
calculated in different FORTRAN codes
and methods are presented on Fig.~\ref{du}. It should be noted
that the lowest order RC are the same for
POLRAD, RADGEN and ESFRAD. But the higher order corrections in these
codes are estimated in different way:
\begin{itemize}
\item
In RADGEN the soft photon approximation is used to account multi--photon
events.
\item
In POLRAD only the second order leading correction is included.
\item
In ESFRAD together with the leading contribution the next-to-leading
one is partly incorporated in accordance with Eq.(3).
\end{itemize}

 It can be seen from Fig. 2(a) that at HERMES kinematics POLRAD
and ESFRAD give very similar results for unpolarized part of
the cross section while the behavior of the higher order
correction generated by RADGEN coincides with
POLRAD and ESFRAD only at large values of the variable $y.$
 At the same time at JLab kinematics (see Fig. 2(b)) the higher
order effects in the considered different approaches
show more distinctions. In this
case the results of POLRAD and ESFRAD agree only at low $y$ and there is
noticeable disagreement between them when $y$ increases and approaches
its maximum value.
On the other hand, RADGEN disagrees with POLRAD and ESFRAD at low $y$ and
coincides with POLRAD at large $y.$
Note that for any given kinematical point
$(x,y)$ the value of RC for JLab is shifted
toward positive values compared to
 the corresponding RC for HERMES.

Radiative corrections to the polarized asymmetries (9), (10) are
presented in Fig. 3 and Fig. 4, respectively. One can see that
the higher order effects
give a very small contribution to the total correction in the entire kinematic
region. It means that the effective higher order correction has a rather
factorized form
and disappears in the ratio of the polarization--dependent and
unpolarized parts of the cross section.
It should be noted that the main difference in the behavior of RC
to $A_1$ and $A_2$: when  $x$ is growing RC to $A_1$ increases
while RC to $A_2$ decreases. As well as for
unpolarized case the value of RC
to asymmetries for JLab is larger than the corresponding RC for HERMES.

\section{Conclusion}

In this report the electron structure function method for
calculation of the higher order radiative corrections
is overviewed briefly
and a comparative analysis of the FORTRAN code ESFRAD developed on the basis of
ESF approach, with POLRAD and RADGEN is performed. All these codes
account exactly for the first--order correction but the higher order effects are incorporated
into them in different ways.

Our numerical analysis of the higher order RC in deep-inelastic events shown that

\begin{itemize}
\item
RC to the unpolarized cross section and the spin asymmetries at JLab kinematics
conditions ($E_{beam}=4$ GeV) are larger than at HERMES ones
($E_{beam}=27.5$ GeV).
\item
The higher order corrections generated by ESFRAD and POLRAD have the similar
behavior at HERMES but differ noticeably at JLab conditions.
\item
At large values of $y$ where RC is very important all three codes
give the same result at HERMES kinematics but at JLab the higher order
contribution of ESFRAD decreases considerably the total RC as compared with
POLRAD and RADGEN.
\item
The higher order corrections contribute very little to polarization asymmetries
that means that effectively they have rather factorized form and
practically are
canceled in the ratio of unpolarized and polarized parts of the
cross section
in the entire kinematic region allowed by HERMES and JLab conditions.
\item
With the increasing $x$ the total
RC to asymmetry $A_1$ increases,  while RC to asymmetry $A_2$ decreases
at both HERMES and JLab experimental conditions.
\end{itemize}
\begin{figure}[t!]
\unitlength 1mm
\vspace*{1cm}
\begin{tabular}{cc}
\begin{picture}(60,60)
\put(-10,-10){
\epsfxsize=8cm
\epsfysize=5cm
\epsfbox{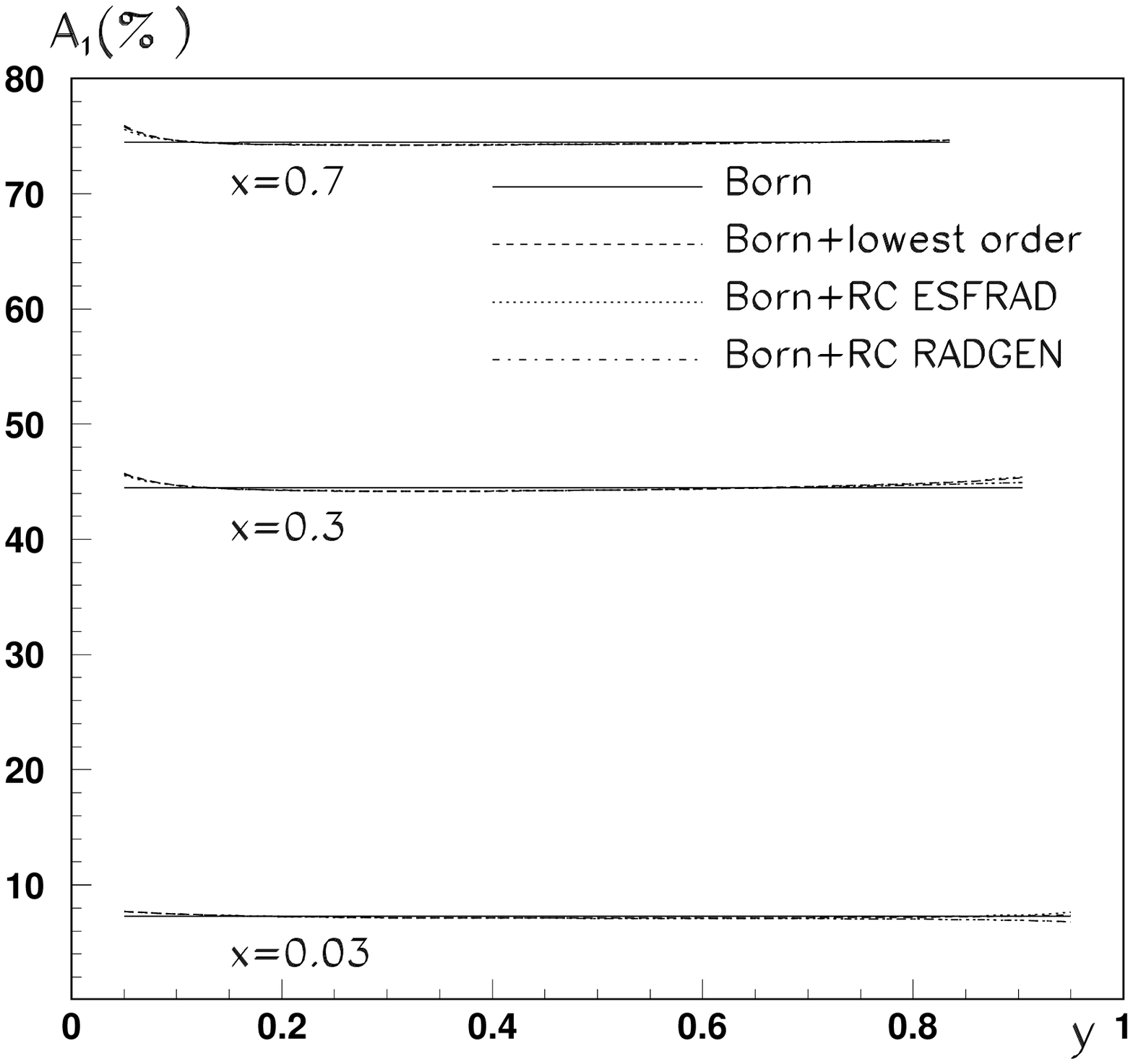}
}
\put(30,-3) {\makebox(0,0){$a)$}}
\end{picture}
&
\begin{picture}(60,60)
\put(10,-10){
\epsfxsize=8cm
\epsfysize=5cm
\epsfbox{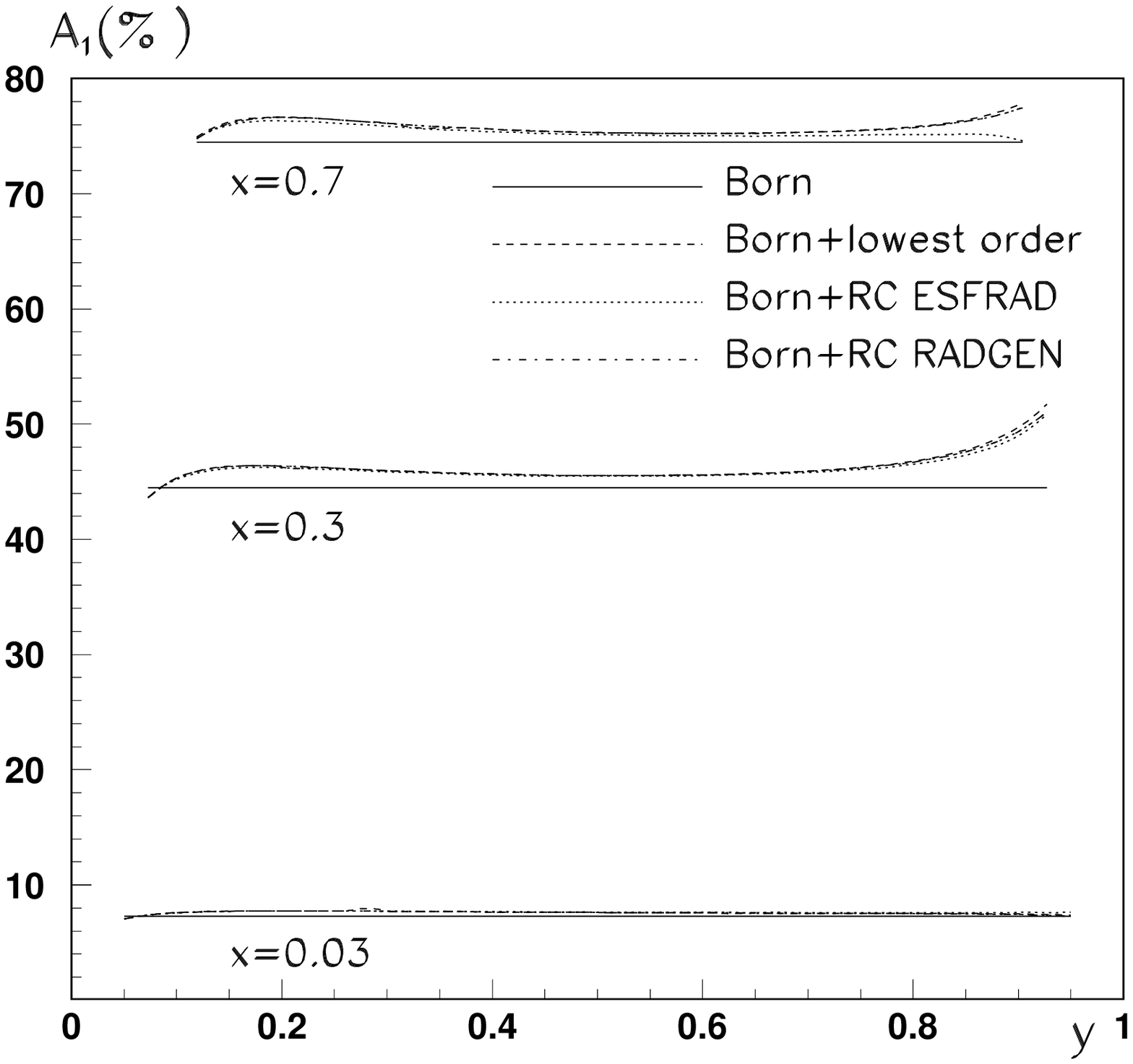}
}
\put(50,-3) {\makebox(0,0){$b)$}}
\end{picture}
\end{tabular}
\vspace*{5mm}
\caption{Radiative corrections to $A_1$ defined by expression
(\ref{a1}) for a) HERMES ($E_{beam}=27.57$ GeV) and b) JLab ($E_{beam}=4$ GeV)
kinematic conditions}
\label{a1rc}
\end{figure}
\begin{figure}[!]
\unitlength 1mm
\vspace*{-4mm}
\begin{tabular}{cc}
\begin{picture}(60,60)
\put(-10,-15){
\epsfxsize=8cm
\epsfysize=5cm
\epsfbox{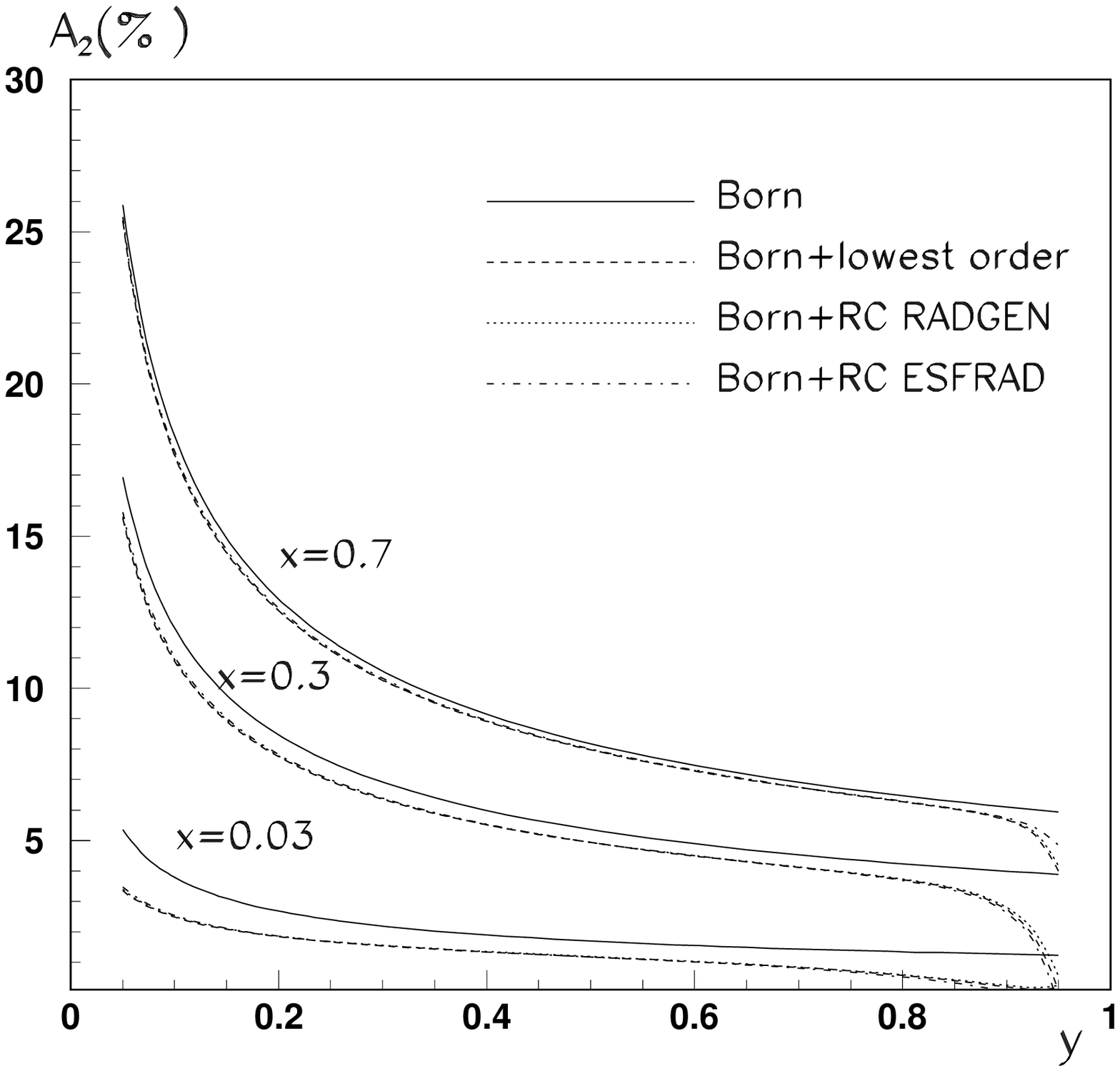}
}
\put(30,-11) {\makebox(0,0){$a)$}}
\end{picture}
&
\begin{picture}(60,60)
\put(10,-15){
\epsfxsize=8cm
\epsfysize=5cm
\epsfbox{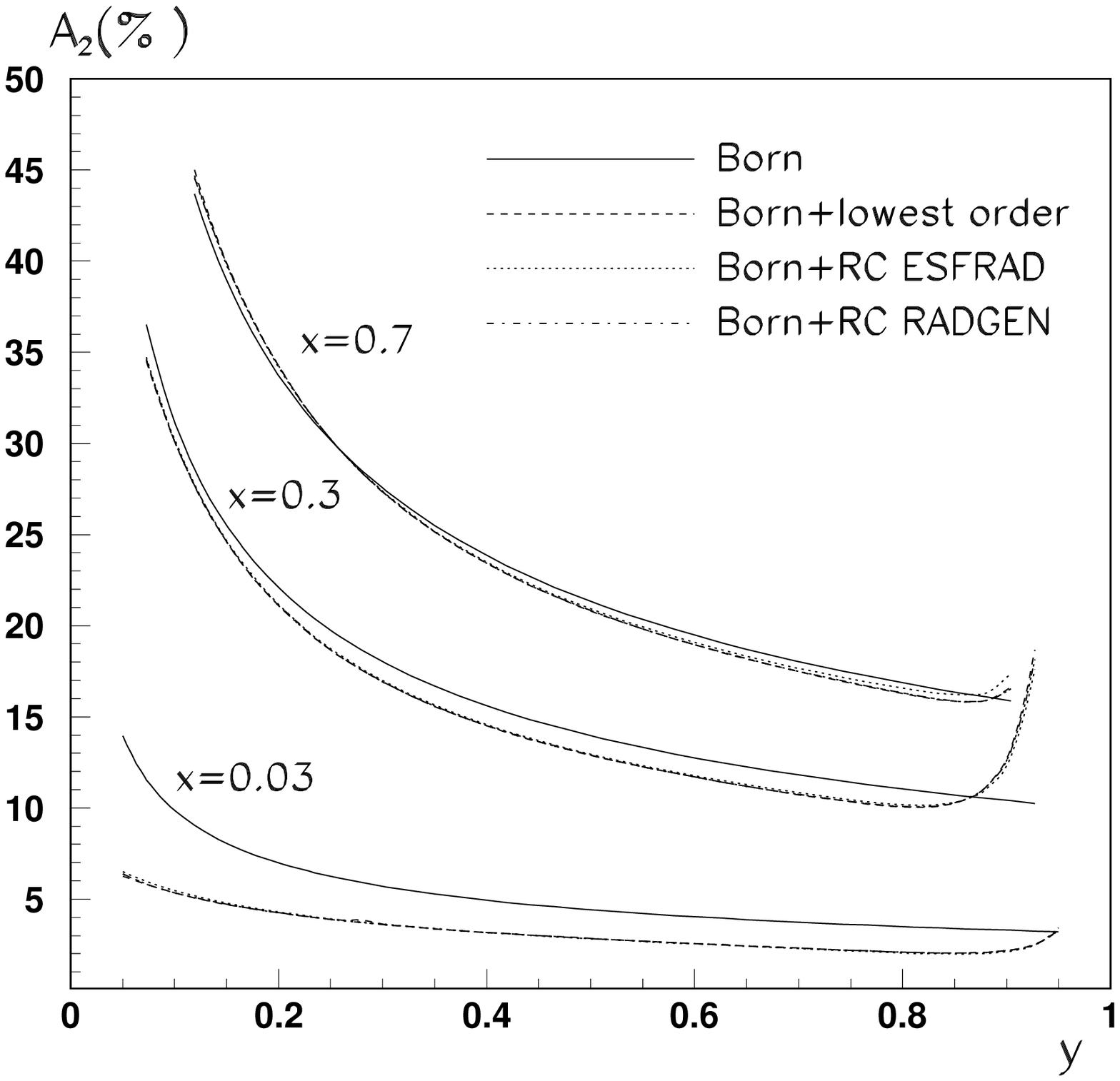}
}
\put(50,-11) {\makebox(0,0){$b)$}}
\end{picture}
\end{tabular}
\vspace*{10mm}
\caption{Radiative corrections to $A_2$ defined by expression
(\ref{a2}) for a) HERMES ($E_{beam}=27.57$ GeV) and b) JLab ($E_{beam}=4$ GeV)
kinematic conditions}
\label{a2rc}
\end{figure}




%

\end{document}